\documentclass[12pt,preprint]{aastex}

\newcommand{\nh}{{N$_2$H$^+$}}
\newcommand{\hcop}{{HCO$^+$}}
\newcommand{\kms}{{km\,s$^{-1}$}}

\slugcomment{To appear in ApJ}

\shorttitle{Global and Local Infall in NGC\,1333}
\shortauthors{Walsh et al.}

\begin{document}


\title{Observations of Global and Local Infall in NGC\,1333}


\author{Andrew J. Walsh, Tyler L. Bourke, Philip C. Myers}
\affil{Harvard-Smithsonian Center for Astrophysics, Mail Stop 42,
60 Garden Street, Cambridge, MA, 02138, USA}
\email{awalsh@cfa.harvard.edu}



\begin{abstract}
We report ``infall asymmetry'' in the HCO$^+$ (1--0) and (3--2) lines 
toward NGC\,1333, 
extended over $\sim 0.39\,{\rm pc}^2$, a larger extent than has been reported before, 
for any star-forming region. The infall asymmetry extends over a 
major portion of the star-forming complex, and is not limited to a 
single protostar, or to a single dense core, or to a single spectral 
line. It seems likely that the infall asymmetry represents inward 
motions, and that these motions are physically associated with the 
complex. Both blue-asymmetric and red-asymmetric lines are seen, 
but in both the (3--2) and (1--0) lines of HCO$^+$ the vast majority of the 
asymmetric lines are blue, indicating inward motions. The (3--2) line, 
tracing denser gas, has the spectra with the strongest asymmetry and 
these spectra are associated with the protostars IRAS 4A and 4B, which
most likely indicates a warm central source is affecting the line profiles. The 
(3--2) and (1--0) lines usually have the same sense of asymmetry in common 
positions, but their profiles differ significantly, and the (1--0) line 
appears to trace motions on much larger spatial scales than does the 
(3--2) line. Line profile models fit the spectra well, but do not 
strongly constrain their parameters. The mass 
accretion rate of the inward motions is of order 10$^{-4}$ M$_\odot$/yr, 
similar to the ratio of stellar mass to cluster age.
\end{abstract}




\section{Introduction}
In the last few years mm-wavelength observations of infall
asymmetry -- the red-shifted self-absorption of optically thick
spectral lines -- have revealed evidence of inward motions in starless
cores, in cores with single embedded stars, and in small cluster-forming
complexes, such as NGC 1333 and Serpens (Myers, Evans \& Ohashi 2000,
and references therein).  
In cores with single stars the infall asymmetry is usually localized 
($<$0.1 pc), and is well matched by the standard model of ``inside-out"
gravitational collapse \citep{shu77} on scales 0.01-0.1\,pc  
(e.g., B335 - Zhou 1995; L1527 - Mardones 1998; Wilner, Myers
\& Mardones 1997; Myers et al.\ 1995; SMM4 - Narayanan et el.\ 2002;
Ward-Thompson \& Buckley 2001; IRAM\,04191 - Belloche et al.\ 2002).  
This is in contrast to starless core maps which show that the infall is much
more extended (0.1-0.3\,pc) than is predicted by the standard model 
(Lee, Myers \& Tafalla 2001).  The cause of these extended motions 
is unclear, and models based on pressure-driven flows, dissipation of
turbulence, supercritical magnetic infall, and early collapse from a condensed
initial state
have been proposed \citep{myers98,nakano98,ciolek00,myers05}.

Such extended infall asymmetry -- global infall -- is also seen in 
limited maps of small-cluster forming regions.
For example, \citet{williams00} report extended infall
asymmetry in Serpens covering a region about 0.2\,pc across, and 
\citet{williams99} surveyed 19 young stellar clusters, but find extended
infall towards only one: Cepheus A.
Also, toward NGC\,1333 IRAS\,4 there
are two main centers of gravity, 4A and 4B, both being multiple
systems. They are separated by 0.05\,pc, but infall
asymmetry is observed over a larger area and encompasses both sources,
with inferred inward speeds of 0.1\,\kms~\citep{mardones98}.
When observed with high resolution (H$_2$CO with 17$''$ -- Mardones
1998; HCO$^+$ with 10$''$ -- Choi et al.\ 1999; H$_2$CO with 2$''$ --
Di Francesco et al.\ 2001), IRAS4 also shows local infall
centered on the protostars,
as indicated by large blue-red ratios in the self-absorbed line profiles,
within 0.02\,pc.
At the finest resolution these appear as inverse P-Cygni profiles.

One explanation for this observation is that the large scale flow
results from a condensation process, while the small scale flow is due to
gravitational collapse.  In one model, the dissipation
of turbulence due to ion-neutral damping of MHD waves, 
leads to an inward pressure gradient and inward flow along magnetic field
lines \citep{myers98}. Alternatively, the motions could be due to gravitational attraction by
an extended layer or network of filaments. Another possibility is that the 
large-scale motions are a poorly-resolved superposition of many localized
collapses.

NGC 1333 is a nearby reflection nebula in the Perseus
molecular cloud complex (e.g., Bachiller \& Cernicharo 1986).
It is a busy region of star formation, containing many Class 0 sources
(e.g., IRAS\,4A, 4B, 2A, 2C and SVS\,13B), as well as more evolved protostars.
Assuming a distance of 300\,pc, using Hipparcos data \citep{dezeeuw99,belikov02}
NGC\,1333 covers approximately a square parsec.
Most of the young stars and protostars appear to be concentrated into a northern
and southern cluster, separated by about 5 arcminutes, equivalent to 0.4\,pc.
Together, the clusters contain at least 143 stars \citep{lada96}.

As part of a survey of nearby clusters and groups, extended infall asymmetry was observed
over a large part of the NGC\,1333 region. In this paper we report on observations
of extended infall in NGC\,1333, in the lines of HCO$^+$ (1--0) and (3--2), and
N$_2$H$^+$ (1--0).

\section{Observations}

\subsection{FCRAO Observations}
Observations with the Five College Radio Astronomy Observatory (FCRAO) were conducted
on 2002, November 16. An area of approximately $16'\times16'$ was mapped in
the lines of \hcop(1--0) and \nh(1--0), covering all of NGC\,1333. The region was
mapped using an on-the-fly technique with
the new 32 element array SEQUOIA, sampling every 0.25 beam, to produce a fully sampled map (FCRAO beam is
52.5\arcsec~at the \nh(1--0) frequency) in 9 hours. The data were regridded onto a
26\arcsec~grid to match up with the CSO observatons detailed below.
The bandwidth was 25\,MHz over 1024 channels, giving a channel spacing of $\sim$0.08\,\kms.
Typical system temperatures were 160
to 180\,K for both transitions. Typically, rms noise estimates are 0.05\,K
per channel in the
center of each map, increasing to 0.08\,K towards the edges of the mapped region.
The rest frequency for \nh(1--0) was chosen to be 93.173258\,GHz for
the `isolated' hyperfine component \citep{lee01}
and the rest frequency for \hcop(1--0) was chosen to be 89.188526\,GHz \citep{pickett98}.

\subsection{CSO Observations}
Observations with the Caltech Submillimeter Observatory (CSO) were conducted 
from 2003, September 14 to 18. Maps of the southern half of NGC\,1333,
including SVS 13, IRAS 4 and IRAS 2, were made in \hcop(3--2). All maps were made
on a 26\arcsec~grid which is approximately beam spaced. Each pointing was
observed with a 50\,MHz bandwidth across 1024 channels, giving
$\sim$0.05\,\kms~channel spacing, and was observed in a
chopping mode with 10 minutes on/off source time.
The assumed rest frequency of \hcop(3--2) is 267.55762\,GHz \citep{pickett98} and
typical system temperatures were 440 to 530\,K.

\section{Results}
\begin{figure}
\plotone{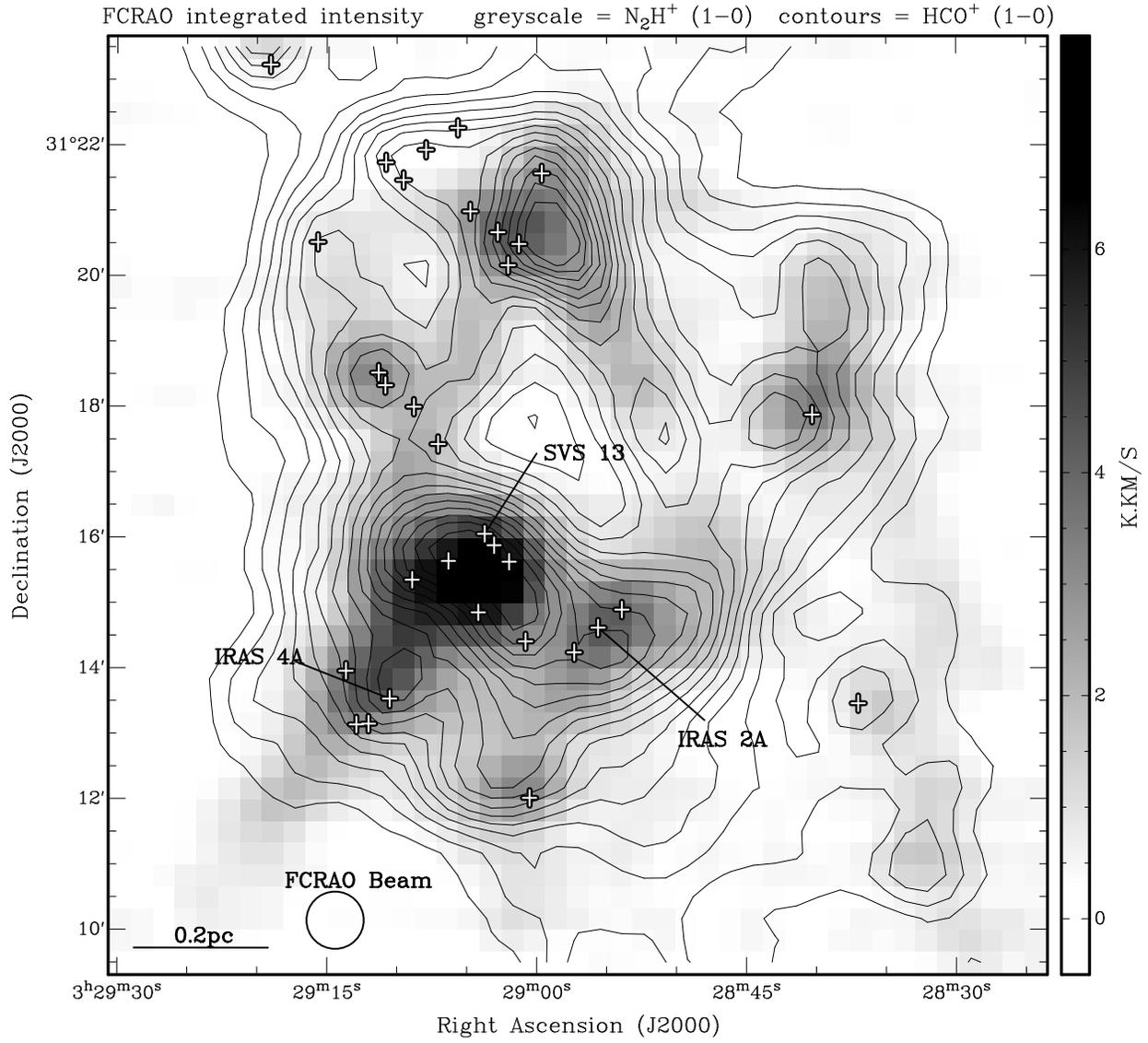}
\caption{Map of NGC\,1333. Greyscale is \nh(1--0) integrated intensity and contours are \hcop(1--0) taken
with the FCRAO telescope, on a 20\arcsec~grid. \hcop(1--0) contours are 10, 15, 20, 25... 95\% of the peak
5.4\,K\,\kms. The plus symbols represent the positions of dust continuum peaks found
by \citet{sandell01}.}
\label{fig1}
\end{figure}

\begin{figure}
\plotone{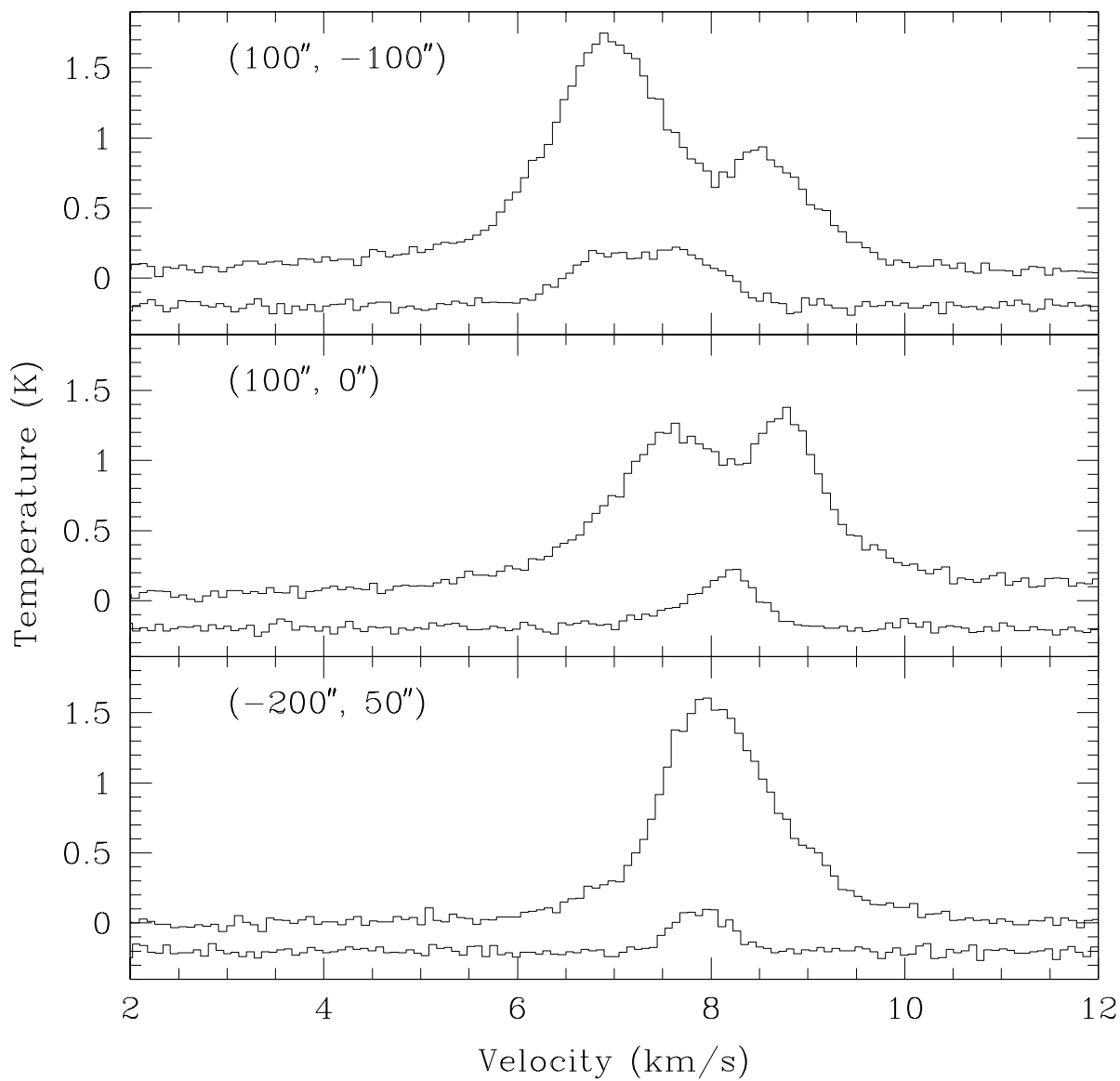}
\caption{Example spectra of HCO$^+$ (1--0) and N$_2$H$^+$ (1--0). For each panel, the top spectrum is HCO$^+$ (1--0)
and the bottom spectrum is the isolated hyperfine line of N$_2$H$^+$ (1--0). The top panel shows an
example of inward motions, the center panel show an example of outward motions and the bottom
panel shows an example where no bulk motions are seen. The (0\arcsec, 0\arcsec) position corresponds to 03 29 03.9 +31 15 18.9
(J2000).}
\label{fig2b}
\end{figure}

\begin{figure}
\plotone{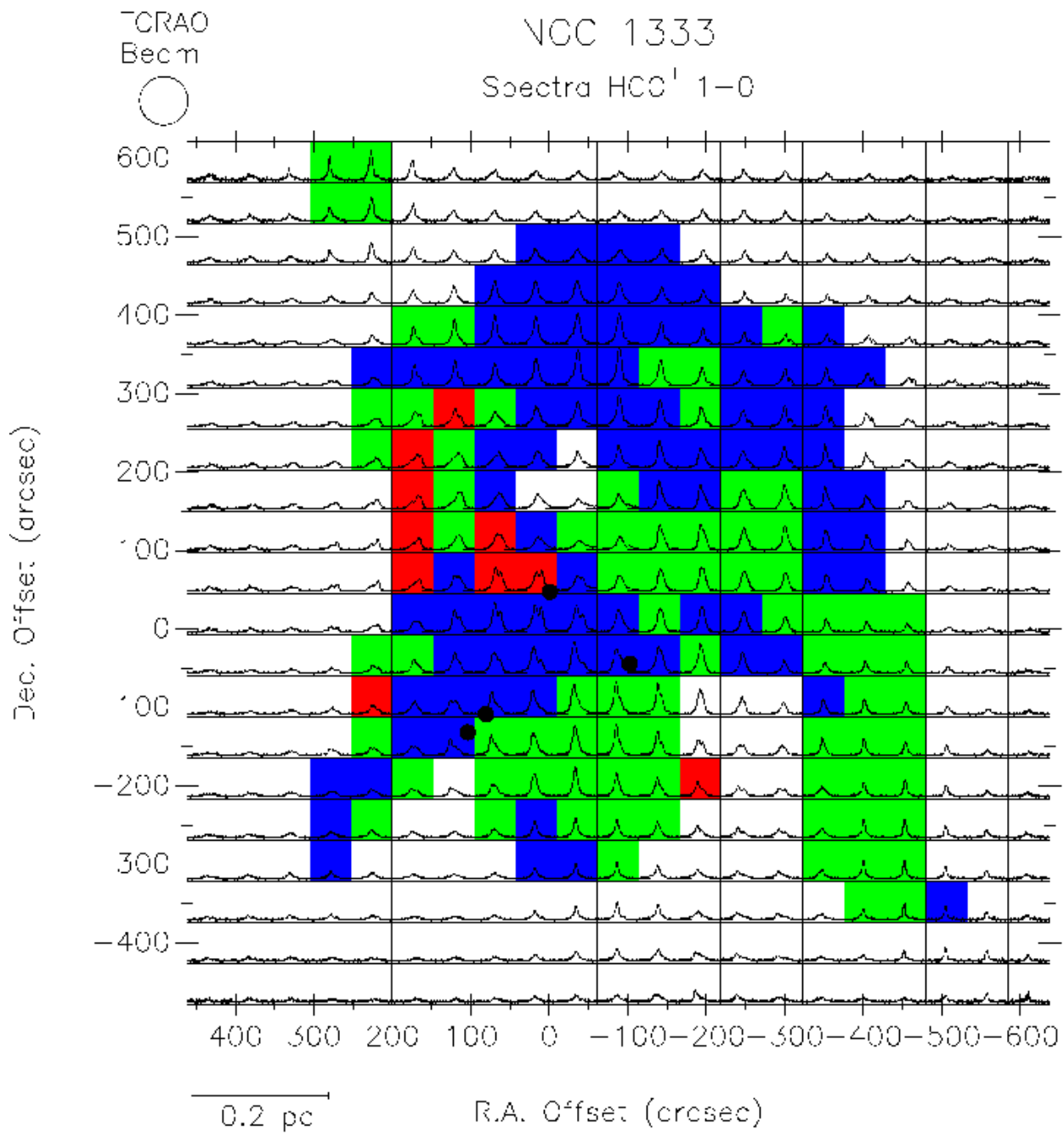}
\caption{Color-coded HCO$^+$ (1--0) spectra. Blue spectra indicate regions with $\delta$v $<$ -0.25 (inward
motions) and red spectra indicate regions with $\delta$v $>$ 0.25 (outward motions). Green spectra
indicate regions where -0.25 $<$ $\delta$v $<$ 0.25 (no clearly defined inward or outward motions).
Uncolored spectra indicate regions where the signal-to-noise is too poor to determine
$\delta$v. The four dots mark the positions of the four main
centers of star formation in the southern half of NGC\,1333; from north to south: SVS\,13, IRAS\,2A, IRAS\,4A
and IRAS\,4B. The scales for the HCO$^+$ spectra are from 3 to 12\,km/s and from -0.2 to 2.5\,K
(${\rm T}_{\rm mb}$). The (0\arcsec, 0\arcsec) position corresponds to 03 29 03.9 +31 15 18.9 (J2000).}
\label{fig2a}
\end{figure}

The \hcop(1--0) and \nh(1--0) FCRAO data are summarised in Figure \ref{fig1}.
In all subsequent figures and discussion, offsets are relative to 03 29
03.9 +31 15 18.9 (J2000).
Figure \ref{fig2b} shows three example spectra, including the optically
thin line \nh(1--0), showing blue-shifted asymmetry, red-shifted asymmetry
and no clear asymmetry.
From an inspection of the \hcop(1--0) spectra in Figure \ref{fig1}, it is clear that many show
blue-shifted asymmetric profiles (``infall asymmetry''). Furthermore, the
optically thin \nh(1--0) emission maximum lies between the two \hcop(1--0)
peaks, indicating that the \hcop~profiles are likely
due to inwardly moving gas that is self-absorbed.
To show this more clearly, we color-code
the \hcop~spectra in Figure \ref{fig2a} where blue and red spectra indicate
blue- and red-shifted asymmetry, respectively. 
We determine the color of each spectrum by using the method pioneered by
\citet{mardones97}, with $\delta$v defined as $({\rm V}_{\rm thick} - {\rm V}_{\rm thin})/\Delta{\rm V}$,
where ${\rm V}_{\rm thick}$ is the peak velocity of the optically thick line (\hcop),
${\rm V}_{\rm thin}$ is the peak velocity of the optically thin line (\nh) and $\Delta{\rm V}$
is the FWHM of the optically thin line.
When $\delta{\rm v} > 0.25$ we assign the spectrum to be red, when $\delta{\rm v} < -0.25$
we assign the spectrum to be blue and when $-0.25 < \delta{\rm v} < 0.25$, we assign the
spectrum to be green.

The blue-shifted asymmetry is seen covering a large area ($\sim0.3$\,pc\,$\times 0.2$\,pc),
including most of the area between the star-forming regions
IRAS\,4, SVS\,13 and IRAS\,2. Towards SVS\,13, the asymmetry appears to reverse,
suggesting significant outward motions in this region, perhaps as a result of the
well studied outflows \citep{knee00}. To the north-west
of NGC\,1333, a second area of extended blue-shifted asymmetry is seen, covering
an area of $\sim0.2$\,pc$\times 0.2$\,pc, near the source IRAS\,5. The two areas join close to SVS\,13
and together cover most of a region 0.85\,pc $\times$ 0.5\,pc, totaling an
area of $0.39\,{\rm pc}^2$.

\begin{figure}
\plotone{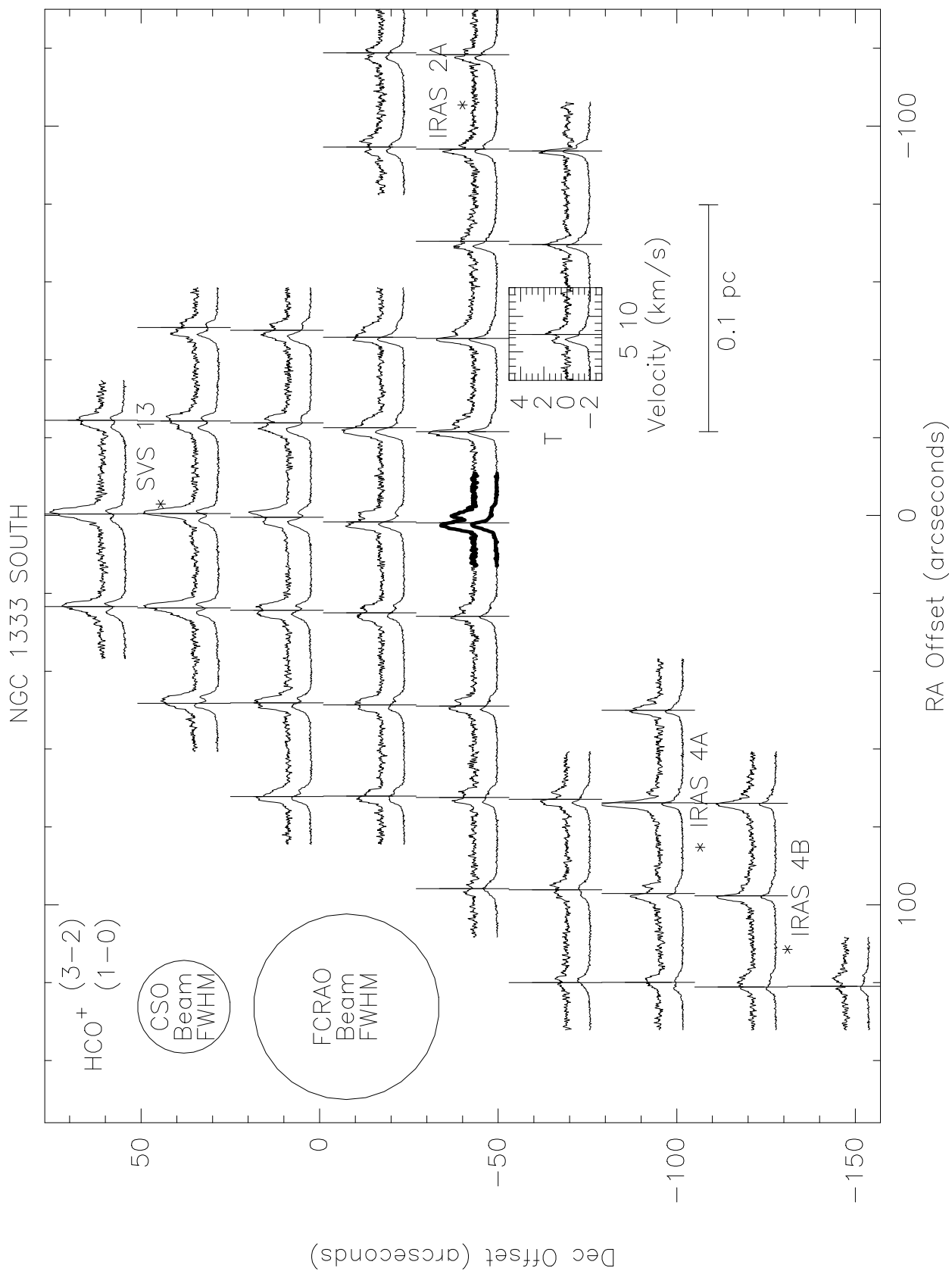}
\caption{HCO$^+$(3--2) (upper) and (1--0) (lower) spectra of the region including IRAS\,4, SVS\,13 and
IRAS\,2 in NGC\,1333. The (0\arcsec, 0\arcsec) position is 03 29 3.9 +31 15 18.9 (J2000). The spectra shown in bold are
located at (0\arcsec, -52\arcsec) and are shown in Figure \ref{fig3}.}
\label{fig2}
\end{figure}

The \hcop(3--2) CSO data are compared to the \hcop(1--0) FCRAO data in Figure
\ref{fig2}. The vertical lines are the systemic velocity determined by fitting
the seven \nh(1--0) hyperfine line components at each position, using
the HFS routine of {\sc class}\footnote{http://www.iram.fr/IRAMFR/GILDAS}. Most of the spectra in this figure show
blue-shifted asymmetric profiles where there appear to be two peaks, the blue being stronger
and the \nh~line located in between the two \hcop~peaks.
Again, around SVS\,13, the asymmetry reverses. Whilst this occurs in both the
(1--0) and (3--2) lines of \hcop, it does not occur at the same positions for
each line, e.g., the (0,0) position. The blue-shifted asymmetry appears to be most
pronounced around IRAS\,4, where inverse P-Cygni profiles have previously been reported
\citep{choi99,difrancesco01}.

\section{Discussion}

\subsection{The Nature of the Extended Infall Asymmetry}
We interpret the blue-shifted asymmetric profiles as infall asymmetry.
The extended areas of infall asymmetry seen in Figures \ref{fig1} and \ref{fig2}
cover much larger areas than previously mapped in regions of isolated star formation
\citep{lee01}. Therefore, a new interpretation of these apparent inward motions
may be required. One possibility is that the single dish observations are of such a low
resolution that we observe a superposition of many sites of local infall over
such a large area. This might be the case because NGC\,1333 is a busy region of
star formation presumably with many sites of localised infall.
However, we consider this an unlikely cause of the observed extended inward motions
for three reasons. First, the
CSO observations are of moderately-high resolution and so are less likely to suffer from
the same confusion problems as the FCRAO observations, and yet we still see similar
infall profiles in both sets of observations covering the same area. Second,
some positions are where
no known star formation is taking place, e.g., 52\arcsec~south of the center position.
Third, to observe infall profiles over such a large area ($>$0.2\,pc) would require a large
number of localized infall regions aligned in such a way as to not leave
any gaps (ie. regions without infall profiles), which would require an
unlikely arrangement.
Therefore we favor the explanation that the infall is physically different from
local infall previously observed, and traces a global inward motion around many sites
of local star formation.

\citet{myers98} suggest that extended infall signatures may be the result of ``turbulent
cooling flows'' associated with dissipation of turbulence.
However, these models are mainly aimed towards explaining extended infall
around isolated low mass star forming regions. One major difference between the theoretical
work and our observations is that the infall speed modeled around isolated low-mass
cores is subsonic, of order 0.1\,\kms, whereas we measure supersonic infall speeds more like 0.5\,\kms~over
extended areas (see \S\S\ref{dmdt}).
It is not clear at this stage whether the models of Myers and Lazarian
can be extended to such high infall speeds.

\citet{walsh04} calculate that speeds of 0.1\,\kms~can be achieved by gravitational
attraction of test particles onto the NGC\,1333 star forming complex in as little as
0.02\,Myr, based on the mass of NGC\,1333: 1450${\rm M}_\odot$, which was derived from
CO measurements \citep{ridge03}. This is assuming that a test particle starts
from rest and falls onto a point mass where gravity is the only force (ie. no pressure
retarding the motion).
Extending this calculation to infall speeds of 0.5\,\kms, we expect speeds of this
order to be reached in 0.1\,Myr, which is about a tenth of the age of the cluster.
This is consistent with global inward
motions of gas falling into the dense centers of NGC\,1333. On the other hand, this model is
extremely idealised and neglects the complex structure and motions of the region.

\subsection{Infall modeling}
\label{infallmodeling}
We have attempted to model physical parameters associated with both the \hcop(1--0) and (3--2)
spectra. To do this, we have chosen a position (located at 0\arcsec, -52\arcsec) that is relatively far
from the main star forming centers, shows a clear infall asymmetry, and exhibits only
minimal outflow wings that may contaminate any modeling. We find that the \hcop(3--2)
spectrum at this position does not show any signs of line wings from an outflow. \hcop(1--0)
does show line wings, which is most likely due to contamination from the IRAS\,2A eastern bow
shock located 49\arcsec~to the west of (0\arcsec, -52\arcsec) \citep{knee00}. We model the
line wings in the \hcop(1--0) spectrum with a single gaussian
having line center velocity 6.9\,\kms, peak intensity 0.11\,K and line width 14.2\,\kms~(FWHM).
\hcop(1--0) and (3--2) spectra at (0\arcsec, -52\arcsec) are shown in Figure \ref{fig3} and the gaussian
modeling of the \hcop(1--0) outflow is shown as the dotted line.

Because there is outflowing and non-outflowing gas present in the spectrum, it is difficult to separate
the two, particularly around 9\,\kms~where both are present in significant proportions. We have
chosen the gaussian that models the outflow as a best guess based on the following criteria:
The \hcop(1--0) spectrum closest to the position of the bow shock (-52\arcsec, -52\arcsec), which shows the strongest
outflow component, is also well modeled by a gaussian with a similar line center velocity
and line width, but stronger peak intensity than the outflow gaussian above, plus two peaks
corresponding to the red and blue shifted lobes of the self-absorbed profile. The outflow contribution
to the (0\arcsec, -52\arcsec) spectrum is unlikely to be the dominant source of emission at 9\,\kms~because we
see emission at this velocity throughout most of Figure \ref{fig2} excluding the
northern part, including positions where there is little or
no redshifted outflow emission. For example, spectra at (26\arcsec, -52\arcsec) to the east and at (26\arcsec, -26\arcsec) to the
north-east of (0\arcsec, -52\arcsec).
Also, Figure \ref{fig3a} shows the \hcop~emission integrated between 8.5\,\kms~and 9.8\,\kms. If we compare this
to Figure 4 of \citet{knee00}, which shows a map of the outflow from IRAS\,2A, we see that our \hcop~emission
does not show the outflow morphology clearly, therefore we do not consider the outflow is the dominant
contributor to the \hcop~emission over this velocity range.

However, while we consider that the outflow contribution at 9\,\kms~is small
compared to the non-outflowing gas, it is possible that the outflow contribution is slightly greater than
we have modeled. We discuss the implications of this below.

\subsubsection{Two-layer model}
\label{2layersect}
\begin{figure}
\plotone{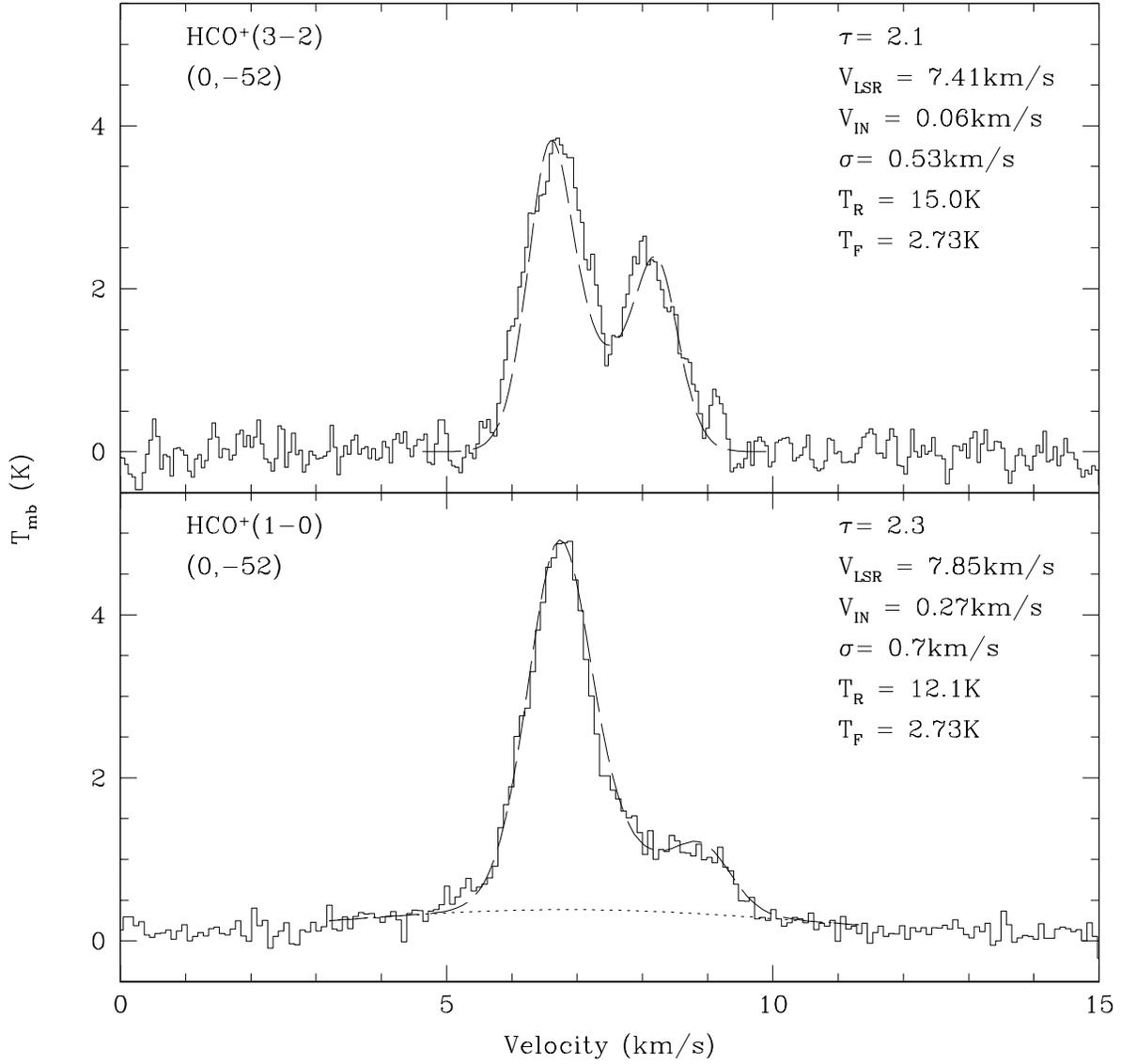}
\caption{HCO$^+$(3--2) and (1--0) spectra of the position (0\arcsec,-52\arcsec) with simple two-layer model
fits shown by the dashed line in each. The dotted line in the 
(1--0) spectrum is a single gaussian fit to the outflow.}
\label{fig3}
\end{figure}

\begin{figure}
\plotone{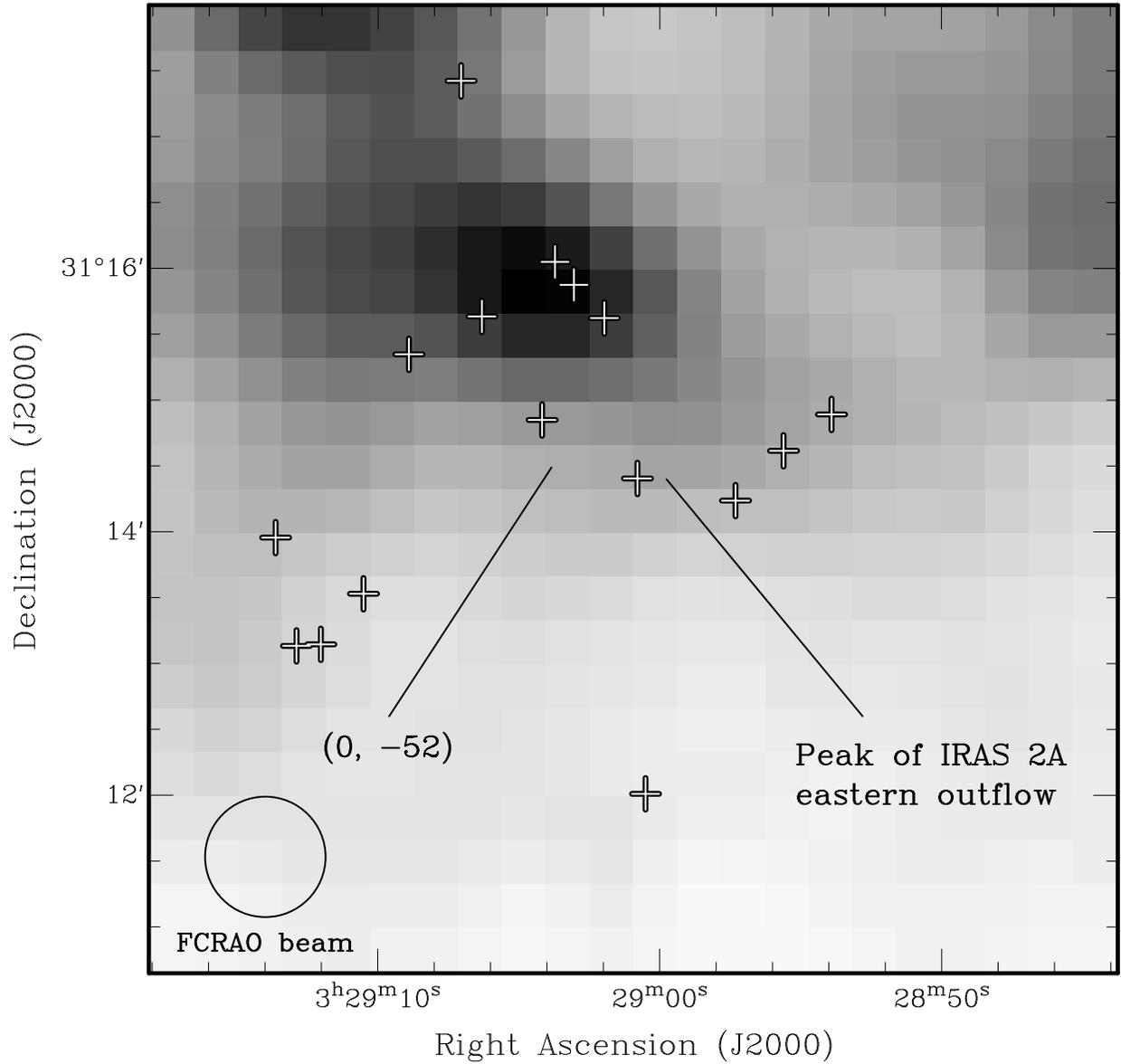}
\caption{HCO$^+$(1--0) integrated intensity map between 8.5\,\kms~and 9.8\,\kms. The position
of the bow shock from the IRAS\,2A eastern outflow is shown, but over this velocity range
the emission is not dominated by the outflow. The plus symbols represent the positions
of dust continuum peaks found by \citet{sandell01}.}
\label{fig3a}
\end{figure}

We fitted each spectrum shown in Figure \ref{fig3}
individually with a simple two-layer model \citep{myers96}. The input
parameters for the two-layer model are the optical depth ($\tau$), systemic velocity
(${\rm V}_{\rm LSR}$), speed at which the two layers are approaching, which is twice the
infall speed (${\rm V}_{\rm IN}$), the velocity dispersion ($\sigma$), the temperature
of the rear layer (${\rm T}_{\rm R}$) and the temperature of the front layer (${\rm T}_{\rm F}$)
which we hold constant at 2.73\,K. This model matches the ``TWOLAYER5'' model of
\citet{devries05}. The best model and parameters
are shown in Figure \ref{fig3}. We find infall speeds and velocity dispersions
of 0.06\,\kms~and 0.53\,\kms~for the (3--2) transition
and 0.27\,\kms~and 0.70\,\kms~for the (1--0) transition of \hcop, respectively. We notice
that some of the parameters for each fit are different: the infall speed is greater for
the (1--0) transition. This is clearly the case as the asymmetry is much greater for the
\hcop(1--0) spectrum. Also, the velocity dispersion is greater for the (1--0) transition,
and the systemic velocity is redder for the (1--0) transition. We note that the systemic velocity
obtained by fitting the optically thin \nh(1--0) line is 7.03\,\kms.
Therefore the systemic velocity determined by the \hcop(3--2) two-layer fit (7.412\,\kms)
is closer to the optically
thin line. A possible explanation for this is that the outflow seen in the \hcop(1--0) spectrum
has not been accurately removed such that a small part remains within the red shoulder of
the \hcop(1--0) spectrum, as discussed above. If we have underestimated the contribution of the outflow,
then this will have the effect of increasing the modeled infall speed of the (1--0) transition. This
is because the weaker red-shifted peak at 9\,\kms~will be reduced by a greater proportion than the
stronger blue-shifted peak at 6.5\,\kms. Even in the extreme case where all of the emission at
9\,\kms~is due to outflow, the blue peak will still be stronger than the (not detected) red peak
of the non-outflowing gas.
Therefore, we find the result that the (1--0) infall speed is higher than the (3--2) infall speed
is robust.

\citet{devries05} also investigate a ``hill'' model, which they show results in a better fit
of infall speeds than this two-layer model. However, rather than including the hill model
here, we use a more detailed radiative transfer model described below.

\subsubsection{Radiative transfer model}
\label{rtsect}
In order to compare the two-layer model with a more comprehensive model, we use the radiative transfer
code {\sc ratran} \citep{hogerheijde00} to simulate emission from the two \hcop~lines.
For our work, we model the emission with 20 concentric shells, each of thickness $10^{14}$\,m.
We constrain the density profile of the emitting region to have a flat inner profile and a profile
falling off as a power law ($\alpha = -2$) of the radius. We also constrain the central
density of the emitting region to be ${\rm n}_c$.
We hold
the radius at which the turnover occurs (5 shells, or $5\times10^{14}$\,m) constant. Such
power-law profiles, with a flat inner region, resemble the structure of a self-gravitating
isothermal sphere, and have been observed by \citet{tafalla04} towards
isolated starless cores. Whilst we do not consider inward motions in NGC\,1333 to be associated
with an isolated starless core, we use this model as a simple approximation.

The simulation allows us to vary infall speed (${\rm V}_{\rm IN}$),
the velocity dispersion ($\sigma$) and the kinetic temperature (T)
for each shell in the emitting region. However, for simplicity, we hold the velocity dispersion 
and kinetic temperature throughout the emitting region to be constant. Furthermore, we restrict
the infall motions to shells 6 to 20 (ie. outside the flat density inner region), which mimicks
infall onto a central static core.
The radial velocity (${\rm V}_{\rm LSR}$) and the relative abundance of HCO$^+$ (X) are free
parameters. We assume a distance of 300\,pc to NGC\,1333 \citep{dezeeuw99,belikov02} and create spectral lines
by integrating over the simulated sphere of emission using the same beam size as the FCRAO and CSO
observations.

\begin{figure}
\plotone{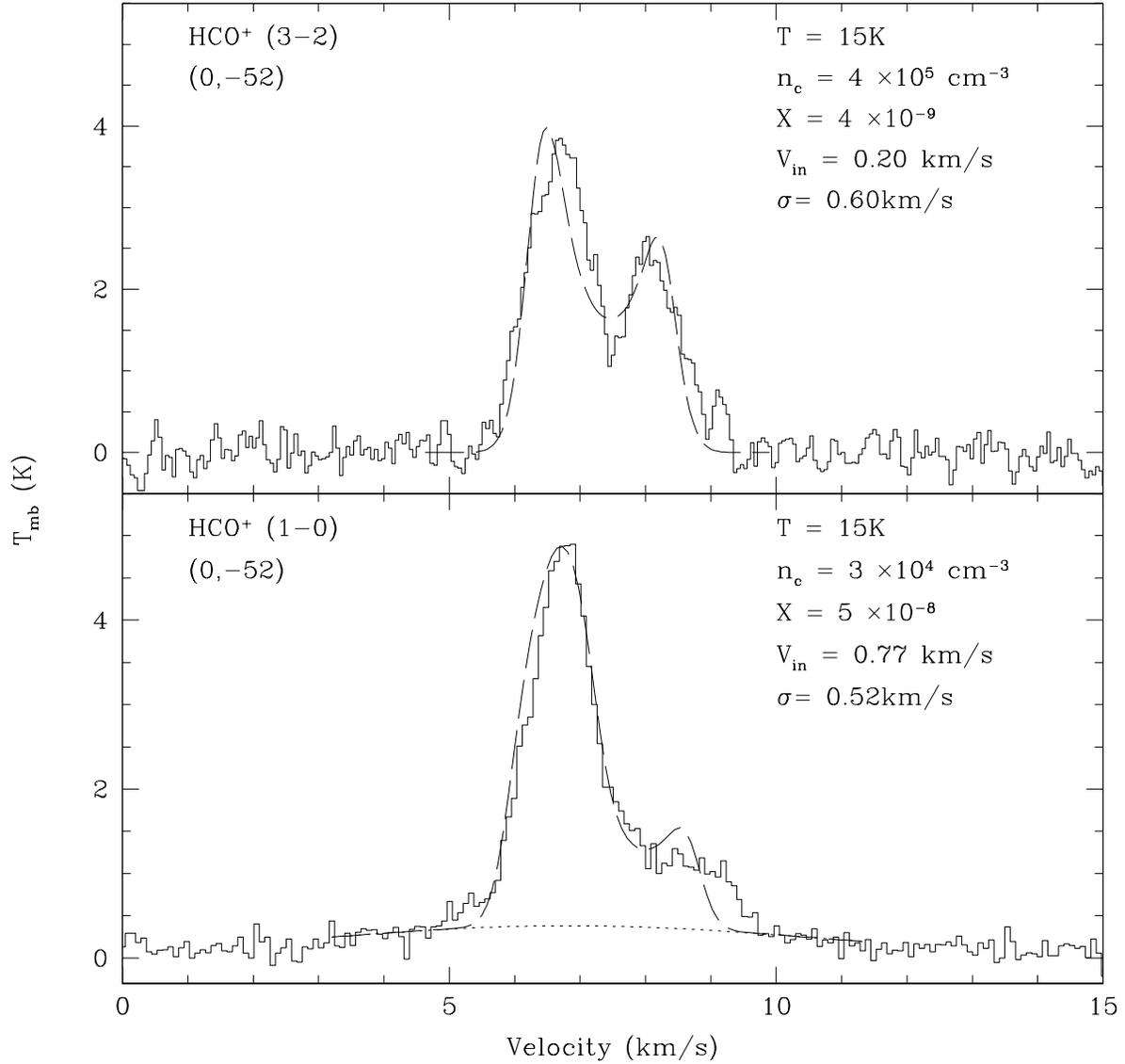}
\caption{HCO$^+$(3--2) and (1--0) spectra of the position (0\arcsec,-52\arcsec) with radiative transfer
({\sc ratran}) fits shown by the dashed line in each. The dotted line in the 
(1--0) spectrum is a single gaussian fit to the outflow.}
\label{fig4}
\end{figure}

Our best radiative transfer fits are shown in Figure \ref{fig4}. The use of this simple model does
not justify a full chi-squared search for the best fitting model. Rather we select the best
fits by visual inspection.
By varying input parameters around the best fit values, we can estimate uncertainties in each
parameter by assessing what appears to be a good fit.
We assign an uncertainty of $\pm 0.03$\,\kms~to ${\rm V}_{\rm IN}$ and $\pm 0.05$\,\kms~to $\sigma$.
However, we caution that not all parameters
are independant: as discussed below, X and n$_c$ appear to be closely linked.
We find that, in the parameter space close to the best fit
values, the infall speed and velocity dispersion are independant of other variables.

The values of the parameters shown in
Figure \ref{fig4} are different for the HCO$^+$\,(1--0) and (3--2) lines. This clearly indicates that
this model cannot fit the two spectra simultaneously, as was found for the two-layer fits.
The reason for this is that the two lines
are most likely probing different regions. The most outstanding discrepancy between the fitted
parameters is for the value of the infall speed. For the (1--0) line it is 0.77\,\kms and for the
(3--2) line it is 0.20\,\kms. This is reflected in the spectra as the asymmetry in the (1--0) line
is much more pronounced than the (3--2) line -- only the infall speed parameter has a major influence
on the asymmetry of the line. As noted in \S\S\ref{2layersect}, if we have underestimated the contribution
of the outflow in the (1--0) spectrum, then this infall speed may be higher. So again we find
that the significant difference between the infall velocities of the (1--0) and (3--2) lines
is a robust result.
The infall speeds determined using the radiative transfer model are higher than
those derived using the two-layer model. We note that
\citet{devries05} have shown that the two-layer model typically underestimates infall speeds by a
factor of two, compared to the radiative transfer model. This is in agreement with our
findings. A possible reason for the lower infall speeds with the two-layer model
is that it does not take into account the portion of the cloud whose systematic line-of-sight
velocity is zero. However, the hill model of \citet{devries05} suffers from the same limitation,
and yet more closely matches the radiative transfer infall speeds. Currently, we are unable to say
what is the reason for the discrepancy between these models.

The best fit values of X and n$_c$ each differ by approximately an order of magnitude between
the (1--0) and (3--2) lines. However, we note that the
simulated spectra obtained by increasing X and decreasing n$_c$ by the same amount have similar
shapes, as long as the change is no more than about an order of magnitude.
As X is increased and n$_c$ is decreased by the same amount, the overall intensity decreases.
The overall intensity is also governed by the kinetic temperature (T), which we have held fixed at
15\,K. But if we allow T to vary, it is possible to
preserve the line shape by, for example, increasing X, decreasing n$_c$ and increasing T.
This implies that, since we do not know the kinetic temperature,
the range of best fit values is an order of magnitude for both X and n$_c$.
The values of $10^4-10^5\,{\rm cm}^{-3}$ for n$_c$ are reasonable considering
\citet{tafalla04} measured values around $10^5\,{\rm cm}^{-3}$ in L1498 and L1517B. The relative
abundance of \hcop~of $10^{-9}-10^{-8}$ agrees with the observationally determined abundance
of $2 \times 10^{-9}$ for IRAS\,16293-2422 \citep{schoeier02}.

\begin{figure}
\plotone{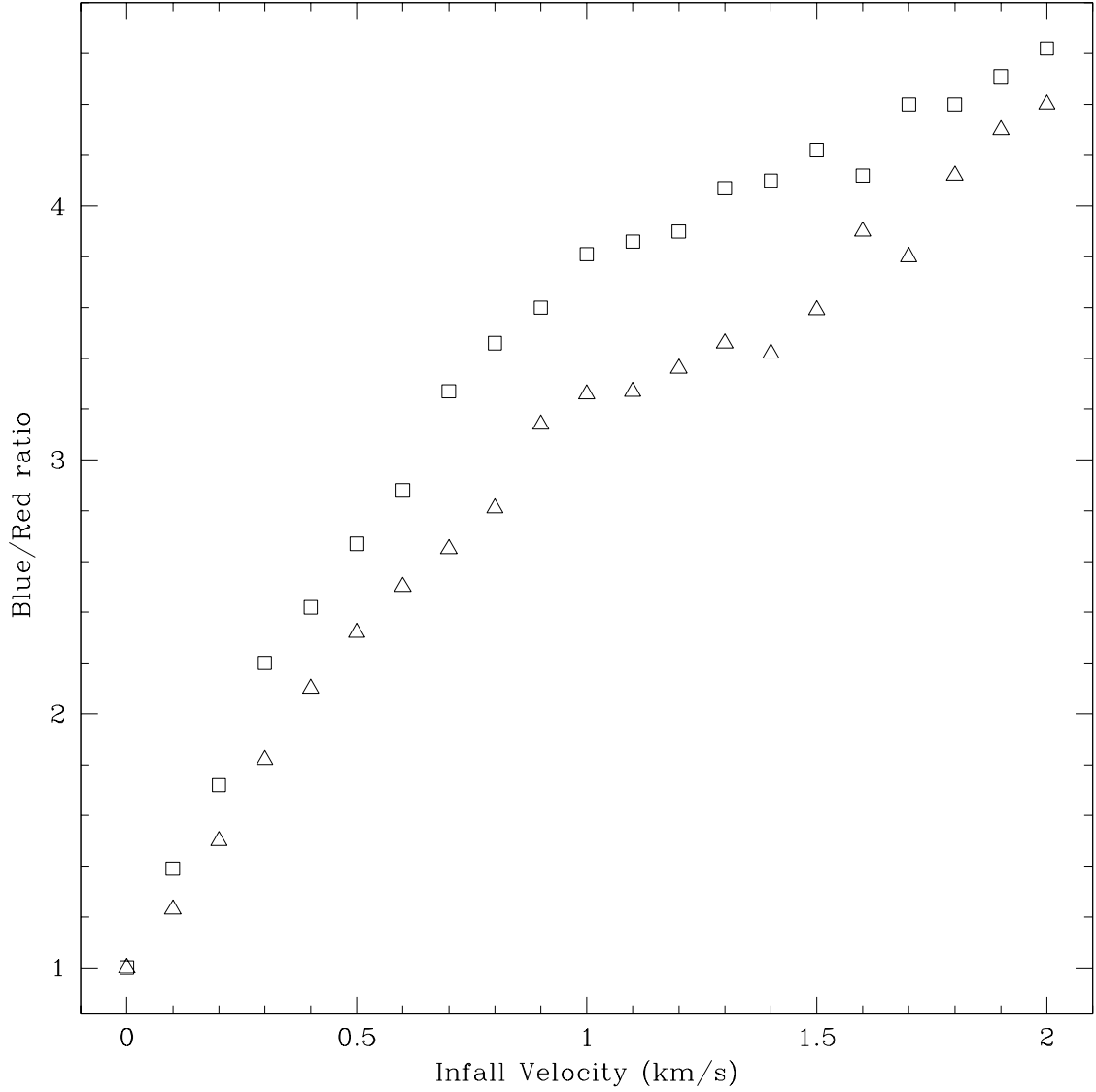}
\caption{Distribution of modeled blue/red peak ratios for various infall velocities. Squares denote
models with other parameters that best fit the HCO$^+$(1--0) spectrum at (0\arcsec, -52\arcsec):
${\rm n}_{\rm c} = 3 \times 10^4\,{\rm cm}^{-3}$, ${\rm X} = 5 \times 10^{-8}$, $\sigma = 0.52\,{\rm km/s}$.
Triangles denote models with other parameters that best fit the HCO$^+$(3--2) spectrum at (0\arcsec, -52\arcsec):
${\rm n}_{\rm c} = 4 \times 10^5\,{\rm cm}^{-3}$, ${\rm X} = 4 \times 10^{-9}$, $\sigma = 0.60\,{\rm km/s}$.}
\label{vin_rbratio}
\end{figure}

Overall, we find that the large number of parameters to fit, as well as the complexity of the best fits
suggests that the radiative transfer model is limited in what it can tell us about the gas
at the position (0\arcsec, -52\arcsec) we have chosen to consider.
However, the spectra shown in Figure \ref{fig4} clearly show a difference in
the degree of asymmetry, indicating a difference in the infall speeds measured by the (1--0) and
(3--2) lines of \hcop. Figure \ref{vin_rbratio} shows the distribution of blue/red peak ratios
as a function of infall speed for \hcop(1--0) and (3--2), holding all other parameters constant at the
best fit values shown in Figure \ref{fig4}. We note that since Figure \ref{fig4} is produced solely from
the radiative transfer model, there is no outflow contamination that might change these results.
The infall speed smoothly increases with increasing
blue/red \hcop~ratio up to infall speeds of about 1\,\kms. At higher infall speeds, the blue/red
ratio (ie. ratio greater than about 3.5) becomes less sensitive to the infall speed. This is expected because
an infall speed of 1\,\kms~is about twice the velocity dispersion of the line (0.52 or
0.60\,\kms~in our models). Therefore, the front infalling material no longer absorbs the emission from the
rear, and the relative intensity of the blue and red components becomes independant of infall speed.
So we can only reliably infer infall speeds when the blue/red ratio is below $\sim$3.5.

\subsection{Infall speed maps}
\begin{figure}
\plotone{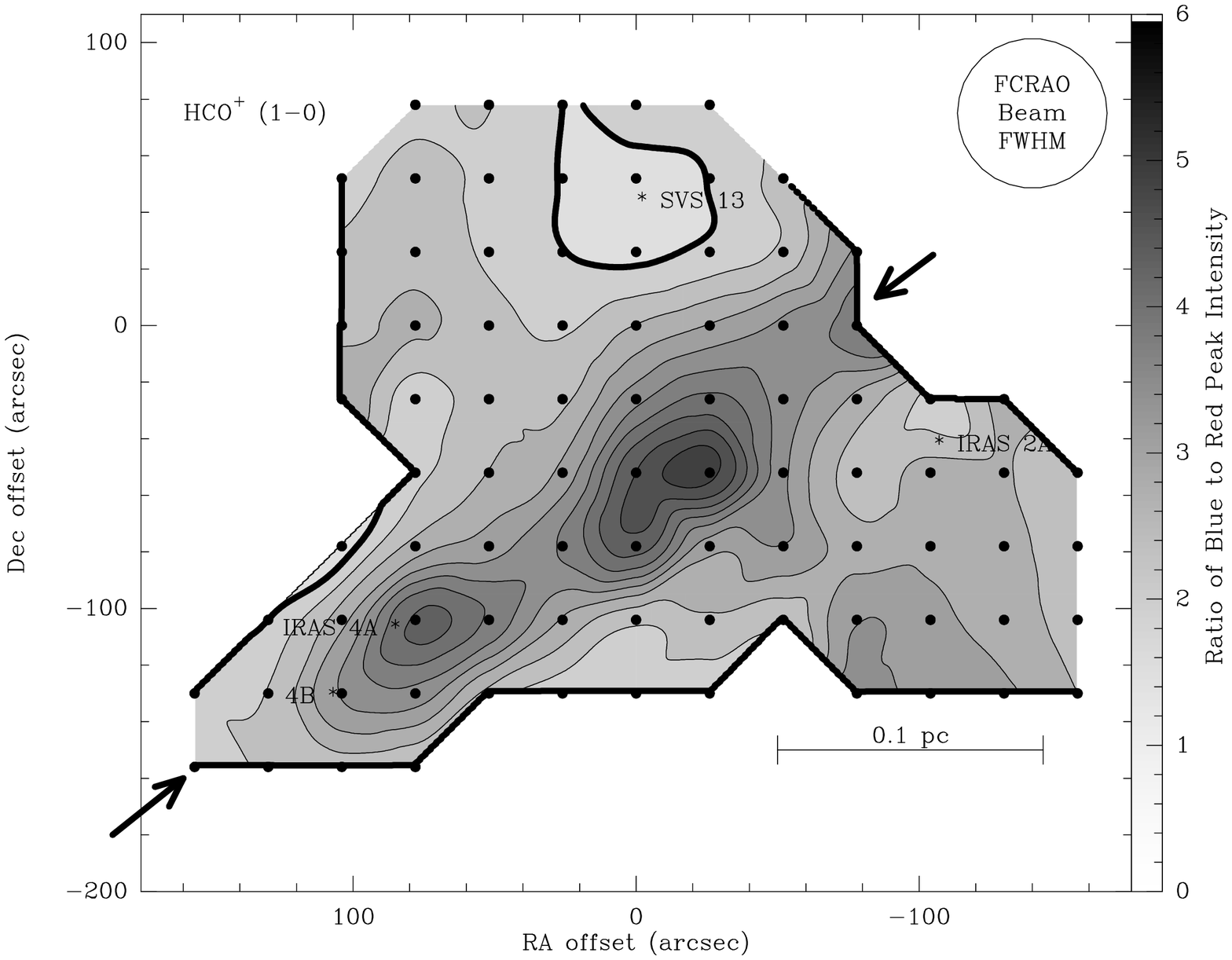}
\caption{HCO$^+$ (1--0) intensity ratio map. The (0\arcsec, 0\arcsec) position is 03 29 3.9 $+$31 15 18.9 (J2000).
The thick lines show the contour where the red and blue lines are the same height, ie. no asymmetry.
The arrows show the orientation of the cut shown in Figure \ref{ratio_cut}.}
\label{fig5}
\end{figure}

\begin{figure}
\plotone{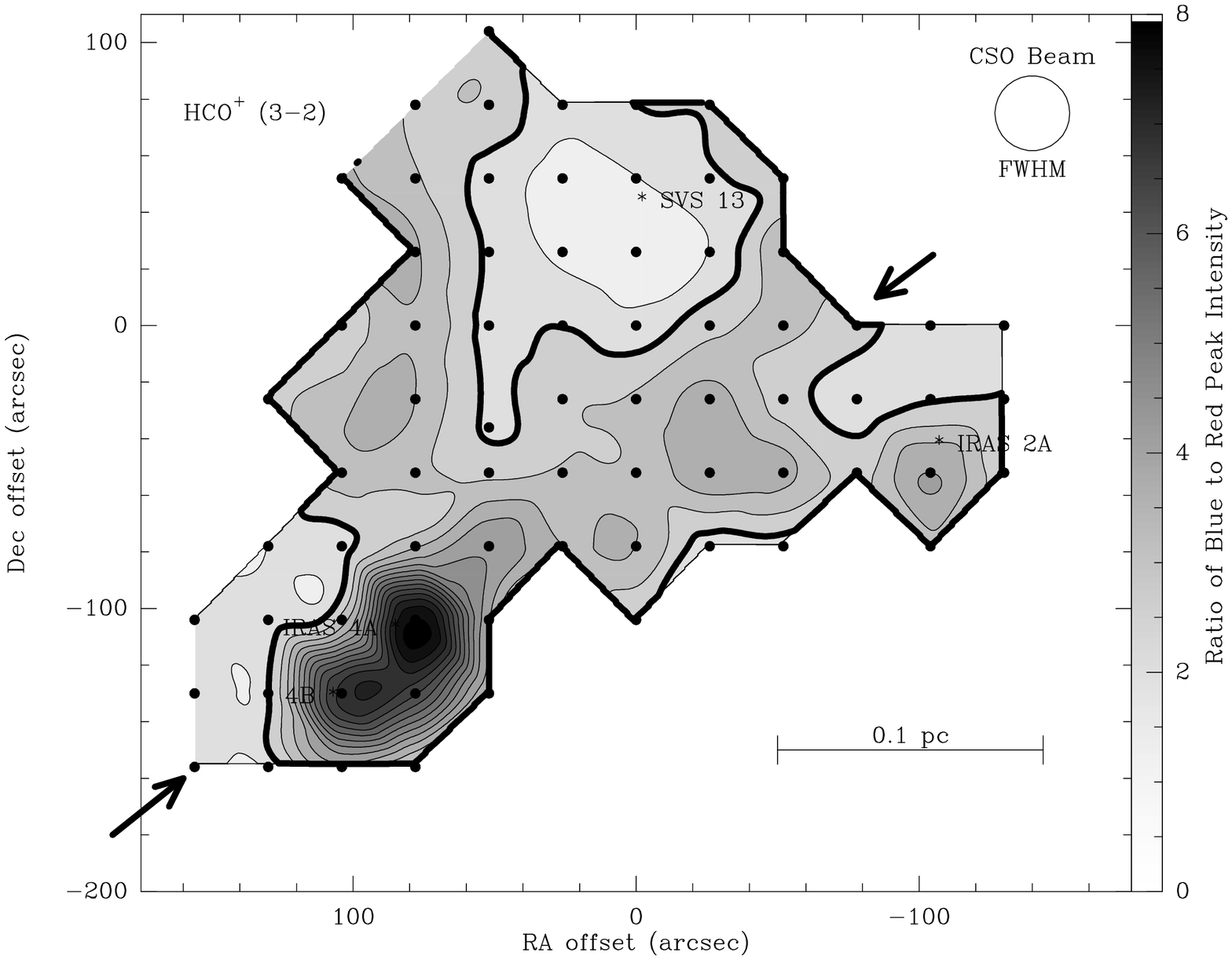}
\caption{HCO$^+$(3--2) intensity ratio map. The (0\arcsec, 0\arcsec) position is 03 29 3.9 $+$31 15 18.9 (J2000).
The thick line shows the contour where the red and blue lines are the same height, ie. no asymmetry.
The arrows show the orientation of the cut shown in Figure \ref{ratio_cut}.}
\label{fig6}
\end{figure}

As can be seen above, the degree of asymmetry of the self absorbed, optically thick line profile
is governed by the infall speed, for sufficiently small blue/red ratios.
Although we cannot be sure about all the best fit
parameters, we can compare relative infall speeds at different positions in the map.
From a comparison of the fits to the spectra in Figures \ref{fig3} and \ref{fig4},
it is difficult to model the emission with simple models. This is mainly because the region is
complicated and a simple spherically symmetric model does not adequately represent the region.
However, our modeling indicates that the ratio of the blue to red peaks in the spectrum is a
good indicator of the infall speed. In both the two-layer and radiative transfer models, only
the infall speed appears to significantly affect the ratio of the red and blue peak heights
provided the infall speed is less than about twice the velocity dispersion, as discussed in
\S\S\ref{rtsect}. All other
variables may have small effects on the blue/red ratio, but they drastically change the line profile at
the same time. Therefore, to preserve a line profile similar to those observed, we assume that only the
infall speed is responsible for changes in the ratio. We present maps of blue to red peak ratio for
HCO$^+$\,(1--0) and (3--2) in Figures \ref{fig5} and \ref{fig6}, respectively. The height of each blue
and red peak is determined by fitting two gaussians to the line profile and using the height of each gaussian.
Where an outflow signature is suspected, a third gaussian was used to remove the outflow contribution,
in a similar manner to that discussed on \S\ref{infallmodeling}.

The thick lines shown in Figures \ref{fig5} and \ref{fig6} show where infalling
motions change to outward motions, particularly around SVS\,13 (ie. where the lines are symmetric).
The extent of outward motions
from both HCO$^+$ transitions is similar but not identical. This region appears to be larger in the (3--2)
transition, compared to the (1--0) transition. This region overlaps with an outflow coming from the SVS\,13
region, therefore it is possible that the outward motions we see here are due to swept up material associated
with the outflow. \citet{zhou95} show that it is possible to have line profiles where the red peak is stronger
than the blue peak with a combination of infall and rotation. However, we do not consider this
to be the case for SVS\,13 because SVS\,13 is presumably the central concentration about which any rotation
is taking place. At the center of rotation, we would only expect to find an infall profile, where the
blue peak is stronger than the red, and yet we find the red peak stronger than the blue at this position.

\begin{figure}
\plotone{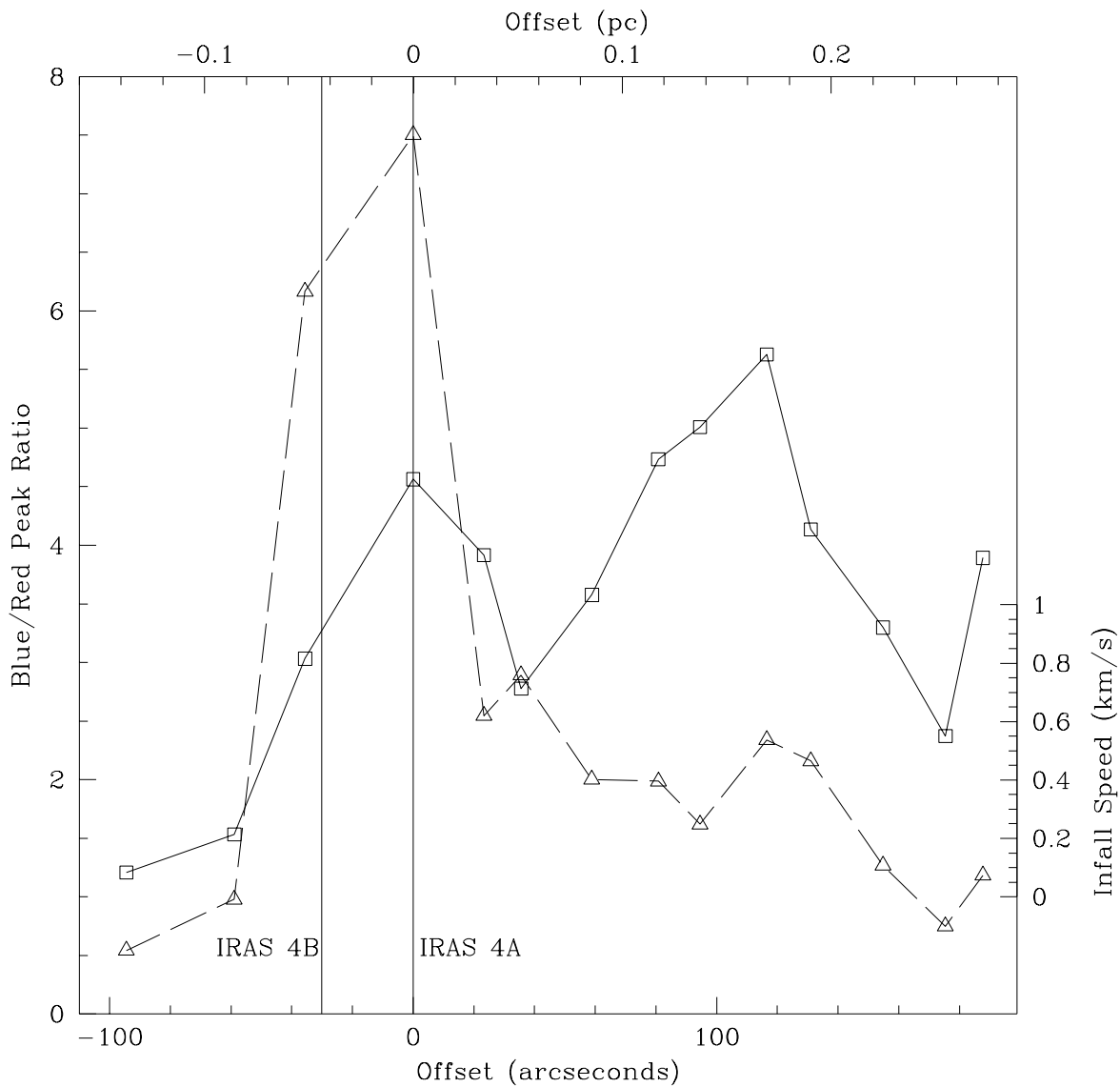}
\caption{Blue/Red HCO$^+$ ratios. Distribution of blue/red peak intensity ratios for HCO$^+$(1--0) is shown
as the squares connected by the solid line, and HCO$^+$(3--2) is shown as the triangles connected by the
dashed line. The positions of IRAS 4A (at the origin) and 4B are marked along the cut. The cut orientation
is shown by the arrows in Figures \ref{fig5} and \ref{fig6}. The right side vertical label shows
infall speeds for the range where the blue/red ratio can be reliably used to estimate the infall speed,
derived from the HCO$^+$ (3--2) transition.}
\label{ratio_cut}
\end{figure}

There is a local maximum of blue/red ratio around
IRAS\,4. This is clearly seen in both transitions, but is more pronounced in HCO$^+$ (3--2),
which also suggests there may be two centers of high blue/red ratio. We speculate that
these two centers may coincide with IRAS\,4A and 4B, but the present data is undersampled
and we cannot claim this conclusively.
This difference in blue/red ratio is more clearly shown in Figure \ref{ratio_cut}, where
the blue/red ratios are shown for a 1-dimensional cut passing through IRAS\,4A and 4B.
In Figure \ref{ratio_cut}, blue/red peak ratios between 2 and 4,
corresponding to infall speeds of 0.4 and 1.0\kms,
appear to be the norm. However, blue/red peak ratios of up to 7.5 occur at the position of IRAS\,4A.
As previously mentioned, our model cannot reproduce such high degrees of asymmetry with high
infall speeds alone. It is also unlikely that such highly asymmetric line profiles are caused by an increase
in optical depth towards IRAS\,4 since the self-absorbed profiles are optically thick; our modeling
suggests line of sight optical depths of 10 are typical. We believe the outflows in the spectra
have been well modeled and are unlikely to be grossly underestimated. In any case, a slight underestimation
of the outflow contribution will have the effect of increasing the blue/red ratio.
We believe the most likely cause of such
highly asymmetric line profiles towards IRAS\,4 is an increase in temperature in the inner part of
the emitting region. This has the effect of increasing the blue peak of the self-absorbed spectrum,
with respect to the red peak. An increase in central temperature is also expected because IRAS\,4A and 4B
both harbor protostars capable of heating up their immediate surroundings.

Presumably the distinction between 4A and 4B is not seen in the (1--0) map due to lower spatial
resolution. If the high blue/red ratios around IRAS\,4A/B occur on small spatial scales, then the
observed increase in ratios, which is greatest for the (3--2) line around 4A/B, 
may be partly explained by the lower resolution of the (1--0) map.
However, the main reason for the higher blue/red ratios in the (3--2) line is almost certainly because it traces
higher density gas than (1--0): we would expect to see the denser gas closer to the protostar(s) and therefore warmer
in the center.

The HCO$^+$ (1--0) map (Figure \ref{fig5}) shows another local maximum of blue/red ratio
centered approximately on the position (-26\arcsec, -52\arcsec), but the (3--2) transition (Figure \ref{fig6}) does not show this.
This is also seen in Figure \ref{ratio_cut}, where an increase in blue/red peak ratio up to 5.5
is seen in \hcop (1--0), but the blue/red peak ratio of \hcop (3--2) remains approximately 
constant at this position.
The nature of this region of increased blue/red ratio is puzzling because it does not occur at any known site
of star formation. However, it is 49\arcsec~to the east of the eastern bow shock of the
outflow from IRAS\,2A \citep{knee00}. We note that we do not consider the increased blue/red ratio
to be an artefact of the outflow because we have removed this component from the spectra. \citet{sandell01}
report the detection of a weak continuum source at this position, who suggest it is a result of the bow shock.
One possible explanation of this is that the bow shock is a site of triggered star formation and we are seeing the
first stages of either increased infall speeds or increased heating as a protostar begins to form.
However, one would expect to
see an even greater blue/red ratio in the (3--2) transition in this scenario, but this is not
the case. In fact, there is no evidence for an increase in blue/red ratio in the (3--2) line at
this position. The outflow is red-shifted on this side of IRAS\,2A, so it is possible that the apparent
increase in blue/red ratio is an artifact produced by increased red-shifted absorption of the line
profile by outflowing material. The lack of evidence for increased (3--2) blue/red ratio is then explained
because the density of material in the outflow is not high enough to absorb significant amounts in the
\hcop(3--2) transition.

It is also intriguing that this region approximately coincides with the position of brightest \nh~emission.
Furthermore, whilst \citet{sandell01} note a weak sub-mm continuum source close to this position (SK14),
it is much weaker than other sub-mm continuum sources in the field (eg. those associated with IRAS\,4A, 4B, 2A and
SVS\,13). Such a discrepancy between the \nh~brightness and the sub-mm continuum brightness suggests this region
may be unusually cold and dense, as \nh~tends to occur in such regions where the main destroyers of \nh~(CO and \hcop)
tend to freeze out onto grains.
In order to reduce the speculation, more observations are needed towards this region.

\subsection{Extended infall or a cold foreground absorbing layer?}
Recently, it has been suggested (Choi et al. 2004; hereafter CKTP04) that the dip in the line profiles
seen towards IRAS 4A/B may be due to an unrelated foreground absorber at a systemic velocity of $\sim$8\,\kms.
CKTP04 suggest that this foreground layer is extended
over a large area, covering at least the regions of IRAS\,4A/B and SVS\,13. CKTP04 favor the interpretation
of a cold foreground layer associated with SVS\,13 because the dip seen in their spectra covers a much larger
area than just IRAS\,4A/B and the emission at the dip velocity appears to peak at the position of SVS\,13.
We favor the interpretation that the line profiles are indicative of inward motions. If there were a cold
foreground layer, then we would expect optically thin \nh~to peak at the same velocity as either the strongest
\hcop~peak or at both \hcop~peaks. However, \nh~peaks between the \hcop~peaks, as would be expected for a
self-absorbed infall profile.
Also, our extended \hcop~maps indicate that the dip covers an even larger area than
previously thought (cf. Figure \ref{fig2a}) and is not centered on SVS\,13. Furthermore, Figures \ref{fig5}
and \ref{fig6} show that the infall speed is most likely the highest at IRAS\,4A.
The absorption dip is not simply deepest at IRAS\,4A, which would be consistent with an unrelated foreground
screen, but also is reddest at IRAS\,4A, which requires a physical association.
Also, the extended infall asymmetry that covers most of the field of view in Figures \ref{fig5} and \ref{fig6}
appears to have a similar
velocity as the inverse P-Cygni profiles detected by \citet{difrancesco01}. We note that \citet{williams00} find
a similar extended infall asymmetry towards Serpens. Therefore we consider the dip to be purely the result
of self-absorption by a large-scale distribution of physically associated gas
and not due to an unrelated foreground layer.

\subsection{Mass infall rates}
\label{dmdt}
Previously, \citet{difrancesco01} calculated the mass infall rate ($\dot{\rm M}_{\rm in}$)
around IRAS 4A and 4B to be $1.1 \times 10^{-4}
\,{\rm M}_\odot\, {\rm yr}^{-1}$ and $3.7 \times 10^{-5}\, {\rm M}_\odot\, {\rm yr}^{-1}$, respectively.
This is based
on the measured infall speed (${\rm V}_{\rm in}$),
the size of the infalling region (${\rm r}_{\rm in}$), the effective critical density of the
transition probing the infall (n):
\begin{equation}
\dot{\rm M}_{\rm in} = 9.6 \times 10^{-5} \left(\frac{{\rm r}_{\rm in}}{0.01\,{\rm pc}}\right)^2 \left(\frac{\rm n}
{1.3 \times 10^6\,{\rm cm}^{-3}}\right) \left(\frac{{\rm V}_{\rm in}}{1.0\,{\rm km}\,{\rm s}^{-1}}\right)
{\rm M}_\odot\,{\rm yr}^{-1}.
\label{eq1}
\end{equation}
Using Equation \ref{eq1} above \citep{difrancesco01}, we can measure the mass
infall rates for the IRAS 4 system using HCO$^+$ (3--2), as well as for the major extended infalling region
encompassing SVS 13, IRAS 4 and IRAS 2 using HCO$^+$ (1--0). Effective critical densities are taken from
\citet{difrancesco01} for H$_2$CO and from \citet{evans99} for the HCO$^+$ transitions, as shown in Table \ref{tabone}.
We calculate infall speeds based on integrated
spectra over the regions of interest and use the {\sc ratran} code, described in \S\S\ref{rtsect}. Our mass infall
rates are summarised in Table \ref{tabone}.

Perhaps the most striking result shown in Table \ref{tabone} is that mass infall rates seem to be about the same
from the very large scales to the very small (between 3.7 to 11 $\times 10^{-5}\,{\rm M}_\odot\,{\rm yr}^{-1}$).
We caution that this should not be intepreted to mean that all the
material infalling on large scales will eventually end up close to IRAS 4A or 4B. This is because we cannot be
certain that the infall rate has remained constant throughout time. On the contrary, it is most likely that
there have been very different infall rates at other times. Therefore, we note that the similarity of the large
scale and small scale infall rates appear to be a coincidence, but this facet of the extended infall in NGC\,1333
should be investigated in more detail.

Whilst there appear to be no large differences in the mass infall rates on different scales, we do see a
difference between the \hcop(1--0) and (3--2) infall speeds on the large scale. Figure \ref{fig5}
shows that the inward motions traced by the (1--0) transition tend to be larger and more widespread than inward
motions traced by the (3--2) transition. This suggests that the gas is slowing down at higher densities. Such a
decrease in inward motions agrees well with models of collapsing layers \citep{myers05} and collapsing Plummer-like
spheres \citep{whitworth01}.

The formation rate of stellar mass in NGC\,1333 can be crudely estimated by assuming that
$\sim 100\,{\rm M}_\odot$ have formed in $\sim 1\,{\rm Myr}$, giving
$\sim 10^{-4}\,{\rm M}_\odot\,{\rm yr}^{-1}$, in rough agreement with the estimates in Table \ref{tabone}.

%
%
\section{Summary and Conclusions}
We have observed NGC\,1333 with the FCRAO in the
gas tracers \hcop(1--0) and \nh(1--0). Our observations cover a large area
approximately 15\arcmin $\times$ 15\arcmin~(1.3\,pc $\times$ 1.3\,pc). We have also observed a subset of this region
with the CSO in \hcop(3--2) in order to trace the higher density gas at a higher
spatial resolution. We identify in both \hcop(1--0) and \hcop(3--2)
large areas of red-shifted self-absorption, indicative of infall, covering up
to $0.39\,{\rm pc}^2$. These large scale inward motions do not appear to be
the result of a superposition of many sites of local infall nor due to absorption of a
foreground layer, and suggest this is truly a large scale inward motion of gas.
Furthermore, the large scale inward motions appear to be qualitatively different from
isolated sites of infall because the global infall speeds are supersonic, whereas
local infall speeds are more typically half the sound speed.

We have modeled the \hcop~red-shifted self-absorption profiles from one position using
a simple two-layer model and a more detailed radiative transfer model.  This position shows no
evidence for outflow in \hcop(3--2), and the outflow seen in the (1--0) spectrum is well modeled
with a single gaussian. Both
models show that the (1--0) and (3--2) emission must come from different regions. This is mainly
because the asymmetry in the two spectra are significantly different, implying significantly 
different infall speeds.
The (1--0) line traces a region of higher infall speed (0.77\,\kms) compared
to the (3--2) line (0.20\,\kms).

We use the ratio of the blue to red peaks in the \hcop~spectra
as an indication of infall speed over the entire region. The relative
infall speed maps show that inward motions dominate the region over approximately the same
area in \hcop(1--0) and (3--2). The blue/red ratio strikingly increases towards the star forming sites
IRAS\,4A and IRAS\,4B, which is most likely due to an internal heating source at those positions.
Outward motions, as evidenced by stronger red peaks than blue peaks, are seen
towards SVS\,13, which could be the signature of entrained material in the outflows from the SVS\,13 region.
We find that the mass infall rates appear to be $\sim 10^{-4}\,{\rm M}_\odot\,{\rm yr}^{-1}$ on scales
from 0.01\,pc to 0.32\,pc, which is similar to the ratio of the estimated mass of stars in NGC\,1333
($\sim 100\,{\rm M}_\odot$) to the cluster age ($\sim$1\,Myr).

\acknowledgments
We would like to thank Susanna Widicus for help with the CSO observing as well as Chris de Vries for
generous help with the radiative transfer modeling. We would like to thank an anonymous referee who greatly
improved the quality of this paper.

\clearpage

\clearpage

\begin{deluxetable}{cccccc}
\tabletypesize{\scriptsize}
\tablecaption{Mass infall rates}
\tablewidth{0pt}
\tablehead{
\colhead{Region}&\colhead{infall}&\colhead{effective critical}&\colhead{infall velocity}&\colhead{radius}&\colhead{mass infall rate}
\\
&\colhead{tracer}& \colhead{density (cm$^{-3}$)} & \colhead{(${\rm km}{\rm s}^{-1}$)} & \colhead{(pc)}&\colhead{(${\rm M}_\odot {\rm yr}^{-1}$)}
}
\startdata
IRAS 4A & H$_2$CO\,($2_{11}-1_{10}$) & $1.3 \times 10^6$ & 0.68 & 0.01 & $1.1 \times 10^{-4}$\\
IRAS 4B & H$_2$CO\,($2_{11}-1_{10}$) & $1.3 \times 10^6$ & 0.47 & 0.01 & $3.7 \times 10^{-5}$\\
IRAS 4A/B complex & HCO$^+$\,(3-2) & $6.3 \times 10^4$ & 0.50 & 0.06 & $7.2 \times 10^{-5}$\\
IRAS 4, SVS 13, IRAS 2 region & HCO$^+$\,(1-0) & $2.4 \times 10^3$ & 0.60 & 0.32 & $1.1 \times 10^{-4}$\\
\enddata
\label{tabone}
\end{deluxetable}

\clearpage

\clearpage 

\clearpage 

\clearpage

\clearpage 

\clearpage 

\clearpage 

\clearpage

\clearpage

\clearpage

\clearpage

%

\end{document}